\documentclass{INTERSPEECH2023}
\usepackage{amsfonts,amssymb,pifont}
\usepackage{multirow,color,cite}

\interspeechcameraready

\title{Semantic VAD: Low-Latency Voice Activity Detection for Speech Interaction}
\name{Mohan Shi$^1$, Yuchun Shu$^2$, Lingyun Zuo$^3$, Qian Chen$^3$, Shiliang Zhang$^3$, Jie Zhang$^1$, Li-Rong Dai$^1$}
\address{
  $^1$NERC-SLIP, University of Science and Technology of China (USTC), China\\
  $^2$Tianjin University, China \\
  $^3$Speech Lab, Alibaba Group, China}
\email{smohan@mail.ustc.edu.cn}

\begin{document}

\maketitle
 
\begin{abstract}
For speech interaction, voice activity detection (VAD) is often used as a front-end. However, traditional VAD algorithms usually need to wait for a continuous tail silence to reach a preset maximum duration before  segmentation, resulting in a large latency that affects user experience. In this paper, we propose a novel semantic VAD for low-latency segmentation. Different from existing methods, a frame-level punctuation prediction task is added to the semantic VAD, and the artificial endpoint is included in the classification category in addition to the often-used speech presence and absence. To enhance the semantic information of the model, we also incorporate an automatic speech recognition (ASR) related semantic loss. Evaluations on an internal dataset show that the proposed method can reduce the average latency by 53.3\% without significant deterioration of character error rate  in the back-end ASR compared to the traditional VAD approach.
\end{abstract}

\section{Introduction}
Speech interaction~\cite{PhilipY87} is one of the most representative applications of artificial intelligence, where long audio recordings need to be cut into short audio segments for downstream tasks, e.g., automatic speech recognition (ASR)~\cite{abs-2111-01690}, speech translation~\cite{Niehues19}, speech emotion recognition~\cite{DrakopoulosPSP19}, etc. For this purpose, a front-end voice activity detection (VAD)~\cite{BishnuSAtal1976APR,salishev2016voice,KraljevskiTB15} is often built to distinguish between speech and non-speech periods, as the former is more related to the interaction goals.

Over the last few decades, many VAD approaches have been proposed. Early VAD methods classify speech and non-speech regions based on energy~\cite{tanyer2000voice,davis2006statistical,woo2000robust,yoo2015formant,8786643}, zero-crossing rate~\cite{JunquaRM91}, and periodicity measure~\cite{Tucker1992VoiceAD}. More recently, with the advance in deep learning, the application of deep neural networks (DNNs) has become dominant in this field~\cite{hughes2013recurrent,kim2016vowel,lee2019spectro,kang2016dnn}, where most aim to improve the classification capability. Also, there are some works on improving noise robustness, e.g., ~\cite{zhang2015boosting,masumura2019context,kim2018voice}.

\begin{figure}[t]
	\centering
	\includegraphics[width=0.9\linewidth]{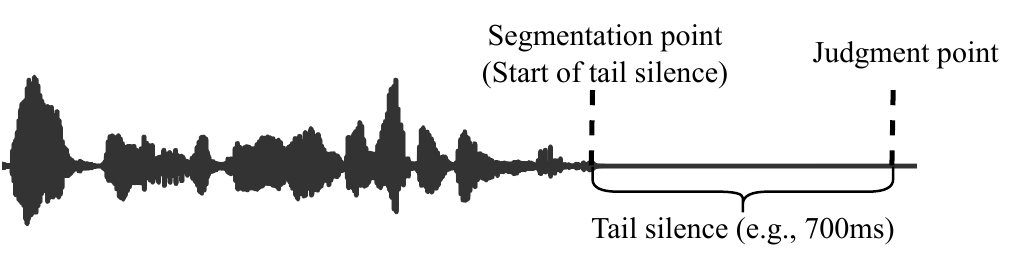}
	\vspace{-0.3cm}
	\caption{
	 Schematic of the tail silence.
	}
	\label{tail silence}
	\vspace{-0.3cm}
\end{figure}

Directly applying these traditional VAD approaches may cause a new problem in realistic speech interaction scenarios. Specifically, traditional VAD algorithms usually need to wait for a continuous tail silence to reach a preset maximum tail silence duration (e.g., 700ms) before deciding whether to perform tail segmentation, e.g., see the example in Figure~\ref{tail silence}, which results in a relatively high tail segmentation latency and seriously affects the user experience~\cite{shangguan2021dissecting,hou2020segment}. In order to address the high-latency issue of tail segmentation, we consider a novel semantic VAD approach in this work. 

Firstly, as traditional VAD models only distinguish speech or silence and neglect whether each part of silence is a complete semantic breakpoint, they have to wait for a long continuous tail silence (e.g., 700ms) before making a tail segmentation judgment to avoid splitting the whole sentences incorrectly. It follows that if a semantic breakpoint is contained within the silence segment, the tail segmentation decision can then be simply made by waiting for a relatively short tail silence. In order to obtain semantic breakpoint information, we thus consider adding a frame-level punctuation prediction task to the basic VAD model. In case an ending punctuation is detected (e.g., period, question mark), indicating that there is a full semantic breakpoint, it is reasonable to wait for a short tail silence (e.g., 300ms). In the case of detecting a non-ending punctuation (e.g., comma), the tail silence required for segmentation needs to be slightly longer than that of ending punctuation (e.g., 400ms). Only when punctuation can not be predicted, the preset maximum tail silence of the traditional VAD (e.g., 700ms) will be used to determine the segmentation point.

Second, in order to further increase the proportion of early judgments and reduce the tail segmentation latency, we add predictions for artificially defined endpoints. Specifically, the endpoint label is introduced in addition to conventional binaural speech-presence and speech-absence categories, forming a three-class classification task. The endpoint label is generated by a combination of punctuation and silence duration, similar to how the punctuation is utilized to determine the tail segmentation. The difference lies in that combining punctuation to determine the tail segmentation is implemented in the inference stage, while combining punctuation to generate endpoint labels artificially changes the training target of VAD.

Finally, since the prediction of punctuation and endpoint requires the model to be semantically strong, we introduce a semantic loss as an auxiliary task to enhance the semantic modeling ability. It should be noted that, different from existing joint VAD and ASR multi-task training methods~\cite{ronny2022e2e,BijwadiaCLSZH22}, where  VAD is cascaded with ASR so that the speech is processed as a whole, resulting in a relatively large parameter amount, while the proposed semantic loss is only used to assist model training and not present at inference time. As such, the independence and lightweight of our semantic VAD model can be guaranteed, meaning that the proposed VAD model can be separately applied or flexibly combined with other downstream tasks.

Evaluations on an internal dataset show that the proposed method can reduce the average tail segmentation latency by 53.3\% without significant deterioration of
character error rate in the back-end ASR compared to the traditional VAD approach. The rest of this paper is orgainzed as follows. Section 2 presents the proposed semantic VAD. Section 3 shows the experimental setup, followed by
results in Section 4. Finally, Section 5 concludes this work.

\section{Semantic VAD}
\begin{figure}[t]
	\centering
	\includegraphics[width=1\linewidth]{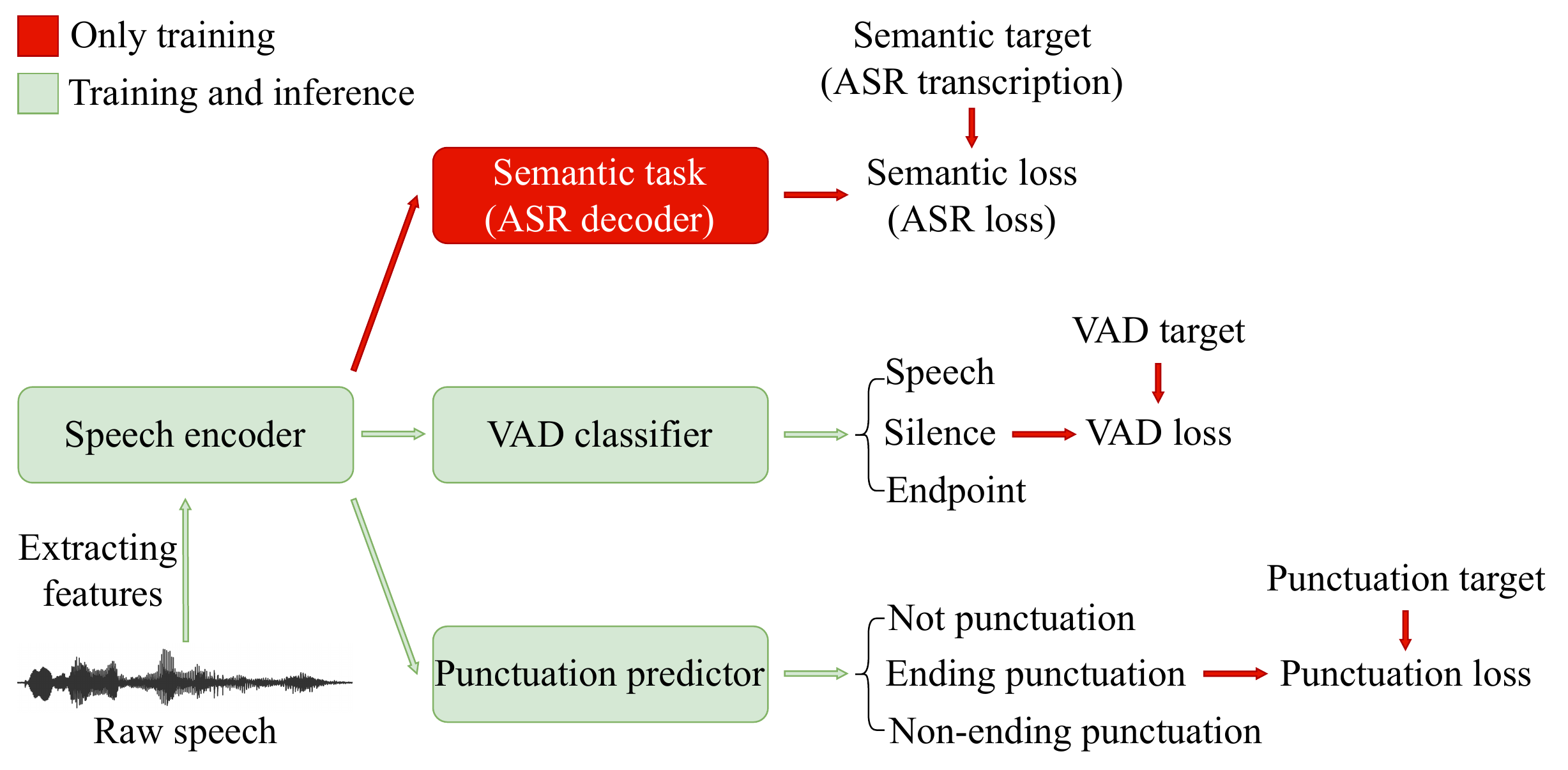}
	\caption{
	 The schematic of the proposed semantic VAD, where the semantic ASR decoder only appears for training  and the rest components are required for both training and inference.
	}
	\label{schematic diagram}
	\vspace{-0.2cm}
\end{figure}

The schematic of semantic VAD is shown in Figure~\ref{schematic diagram}, which consists of the speech encoder, punctuation prediction, VAD classifier, and semantic task. 

\subsection{Speech encoder}

To facilitate the subsequent migration to the streaming platform, we use contextual block conformer~\cite{GulatiQCPZYHWZW20,TsunooKKW19} as our speech encoder, which is a block processing conformer structure designed for streaming ASR. In the speech encoder, a context embedding vector is introduced to convey contextual information between blocks. The acoustic features are passed through the speech encoder to generate a frame-level speech representation.

\subsection{Punctuation prediction}
In traditional VAD, there are no decisions on semantic breakpoints, so it is necessary to wait for a long tail silence before performing tail segmentation. In order to reduce the tail segmentation latency, it might be helpful to obtain semantic segmentation via punctuation prediction. This can be accomplished by directly incorporating the frame-level speech representation obtained by the speech encoder into the classifier to obtain the frame-level punctuation prediction. In this work, we consider three types of punctuations:
\begin{align}
&P=
    \begin{cases}
    \text{Not punctuation} \\
    \text{Ending punctuation (E-punc)} \\
    \text{Non-ending punctuation (NE-punc)}
    \end{cases}
\end{align}
where the ending punctuation means complete semantic breakpoints (e.g., periods, question marks) and the non-ending punctuation implies incomplete semantic breakpoints (e.g., comma).
\subsection{VAD with endpoint}
Conventionally, each speech frame is only simply labeled with  speech-presence or speech-absence. In order to further improve the proportion of tail segmentation points judged in advance and reduce the tail segmentation latency, the artificially defined endpoint label is introduced as the third category, resulting in a three-class classification problem:
\begin{align}
S\in \left\{
    \text{Speech}, 
    \text{Silence}, 
    \text{Endpoint}
\right\},
\end{align}
where the endpoint is the newly defined label in the training set. On the basis of obtaining the labels of speech and silence, in case a sentence ends with an E-punc, the frame after a continuous $t_\text{E}$ silence until the next speech begins will be set with an endpoint; in case of a NE-punc, the duration of the tail silence is then set to be $t_\text{NE}$. 
Similarly to punctuation prediction, the output of the speech encoder has to pass through a classification layer to obtain the frame-level VAD decision.
\subsection{Semantic task}
\begin{figure}[t]
	\centering
	\includegraphics[width=0.9\linewidth]{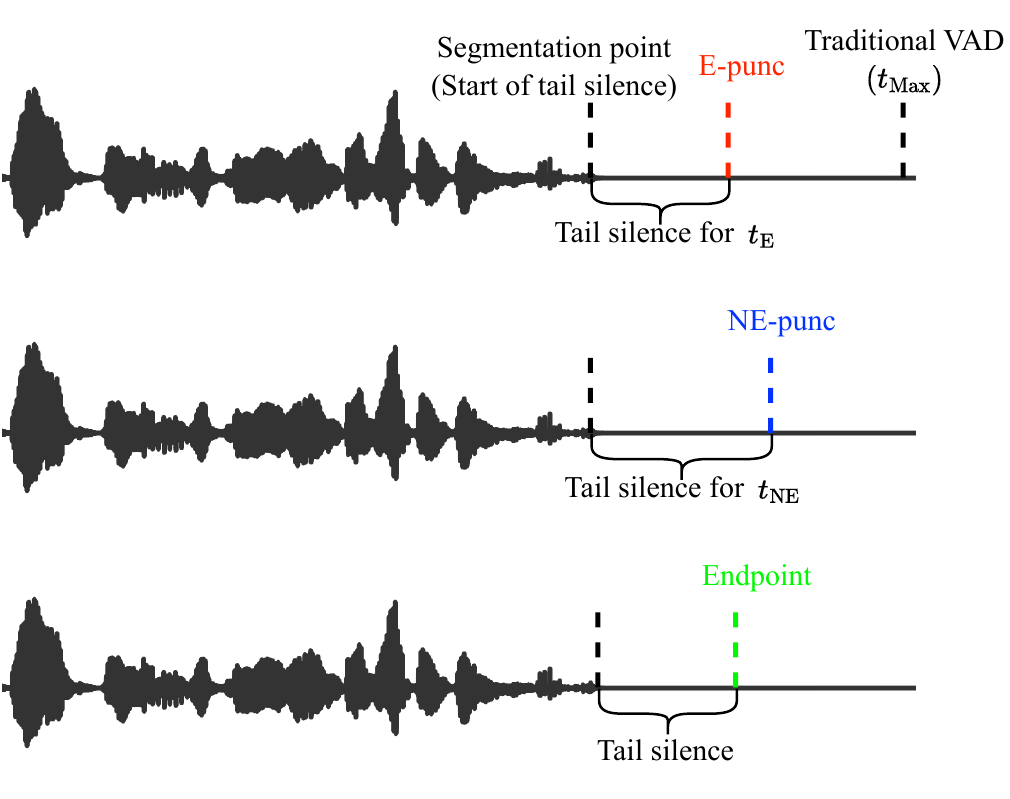}
	\caption{Schematic diagram of several judgment methods during inference.
	}
	\label{inference judgment}
	\vspace{-0.2cm}
\end{figure}
Since both punctuation classification and the prediction of the artificially specified endpoint frames in the considered VAD, the representation obtained by the speech encoder should contain strong semantic information. Therefore, it is necessary to introduce a semantic loss as an auxiliary task to preserve the semantic information in speech representation. In this work, we use the ASR loss as the auxiliary semantic loss. Specifically, the speech encoder is connected to the decoder for pre-training on the ASR task, and the punctuation prediction, VAD and ASR are then jointly trained. The overall loss function for joint training can be represented as
\begin{align}
&L=\mu{L^\text{punc}(\hat{p},p)}+\lambda{L^\text{asr}(\hat{y},y)}+(1-\mu-\lambda){L^\text{vad}(\hat{s},s)}
\end{align}
where $p$, $y$, and $s$ represent the true punctuation, transcription, and VAD label, respectively, while $\hat{p}$, $\hat{y}$, and $\hat{s}$ represent the corresponding hypothesis. Notice that unlike previous multi-task works, ASR is only used here as an auxiliary task to enrich the semantic information during training, and does not appear during inference. This guarantees the lightweight and independence of the front-end VAD module at the inference phase.

\begin{table*}[!t]
\renewcommand {\thetable} {2}
\centering
\caption{Performance comparison of traditional and semantic VAD methods on the three test sets in terms of the tail segmentation latency (ms), CER (\%) for back-end ASR and DCF (\%) for front-end segmentation.}
\setlength{\tabcolsep}{2mm}
\begin{tabular}{c|ccc|ccc|ccc|ccc}
\toprule
\hline
\multirow{2}{*}{Approach} & \multicolumn{3}{c|}{Average} & \multicolumn{3}{c|}{Test1} & \multicolumn{3}{c|}{Test2} & \multicolumn{3}{c}{Test3} \\ \cline{2-4} \cline{5-7} \cline{8-10} \cline{11-13}
\multicolumn{1}{c|}{} & Latency & CER & DCF & Latency & CER & DCF & Latency & CER & DCF & Latency & CER & DCF \\ \hline
\multirow{5}{*}{Traditional VAD} & 700 & 16.75 & 10.78 & 700 & 19.95 & 13.51 & 700 & 27.41 & 9.94 & 700 & 6.78 & 7.60 \\ 
\multicolumn{1}{c|}{} & 600 & 16.80 & 10.78 & 600 & 19.96 & 13.50 & 600 & 27.74 & 9.95 & 600 & 6.77 & 7.61 \\ 
\multicolumn{1}{c|}{} & 500 & 16.98 & 10.70 & 500 & 20.10 & 13.42 & 500 & 28.08 & 9.82 & 500 & 6.93 & 7.55 \\ 
\multicolumn{1}{c|}{} & 400 & 17.42 & 10.71 & 400 & 20.62 & 13.42 & 400 & 28.56 & 9.84 & 400 & 7.23 & 7.56 \\ 
\multicolumn{1}{c|}{} & 330 & 17.94 & 10.70 & 330 & 21.17 & 13.41 & 330 & 29.04 & 9.81 & 330 & 7.73 & 7.56 \\ \hline
Semantic VAD & 326.92 & 16.88 & 9.60 & 326.45 & 19.91 & 11.52 & 338.99 & 27.84 & 9.60 & 326.32 & 7.04 & 7.10 \\ \hline
\bottomrule
\end{tabular}
\label{tab:result_pipline}
\end{table*}
\subsection{Inference}
For traditional VAD methods, after the tail silence reaches $t_\text{Max}$ continuously, it will be assigned a break point, and segmentation is performed at the beginning of the tail silence. This leads to a high tail segmentation latency and thus an annoying user experience. In order to avoid this, we consider four cases to perform tail segmentation: 
\begin{itemize}
  \item An endpoint is detected in the continuous tail silence;
  \item An E-punc is detected and the continuous tail silence reaches $t_\text{E}$;
  \item A NE-punc is detected and the continuous tail silence reaches $t_\text{NE}$;
  \item The continuous tail silence reaches $t_\text{Max}$.
\end{itemize}
These are graphically shown in Figure~\ref{inference judgment}.
It will be experimentally shown that combining the prediction of punctuation and endpoint can greatly reduce the tail segmentation latency.

\begin{table}[!t]
\renewcommand {\thetable} {1}
\centering
\caption{The label distributions (\%) in the training set.}
\setlength{\tabcolsep}{3pt}
\begin{tabular}{ccc|ccc}
\toprule
\hline
\multicolumn{3}{c|}{punctuation labels} & \multicolumn{3}{c}{VAD labels} \\ \cline{1-3} \cline{4-6}
Not punc & E-punc & NE-punc & Speech & Silence & Endpoint \\ \hline
73.79 & 22.05 & 4.16 & 58.51 & 18.80 & 22.69 \\ \hline
\bottomrule
\end{tabular}
\label{tab:label distribution}
\end{table}

\begin{table*}[!t]
\centering
\caption{Performance comparison of the performance of semantic VAD under different strategies on the three test sets.}
\setlength{\tabcolsep}{4pt}
\begin{tabular}{l|ccc|ccc|ccc|ccc}
\toprule
\hline
\multicolumn{1}{c|}{\multirow{2}{*}{Approach}} & \multicolumn{3}{c|}{Average} & \multicolumn{3}{c|}{Test1} & \multicolumn{3}{c|}{Test2} & \multicolumn{3}{c}{Test3} \\  \cline{2-4} \cline{5-7} \cline{8-10} \cline{11-13}
\multicolumn{1}{c|}{} & Latency & CER & DCF & Latency & CER & DCF & Latency & CER & DCF & Latency & CER & DCF \\ \hline
Semantic VAD & 362.95 & 16.92 & 10.91 & 355.29 & 20.08 & 13.47 & 359.07 & 27.32 & 9.43 & 377.20 & 7.13 & 8.23 \\
\quad+Semantic loss & 344.63 & 16.75 & 9.70 & 342.14 & 19.75 & 11.57 & 355.19 & 27.75 & 9.68 & 347.86 & 6.93 & 7.27 \\
\quad\quad+Endpoint & 326.92 & 16.88 & 9.60 & 326.45 & 19.91 & 11.52 & 338.99 & 27.84 & 9.60 & 326.32 & 7.04 & 7.10 \\ \hline
\bottomrule
\end{tabular}
\label{tab:ablation}
\end{table*}

\begin{figure*}[h]
	\centering
	\includegraphics[width=1\linewidth]{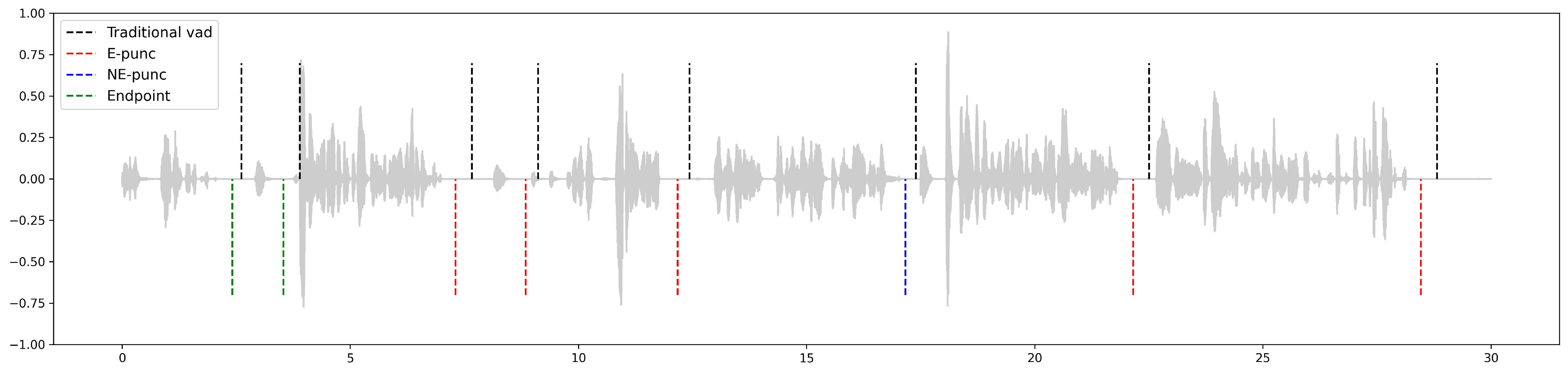}
	\caption{
	 The visual comparison of the tail judgment points of semantic and traditional VAD methods.
	}
	\label{visualize}
\end{figure*}
\begin{figure}[h]
	\centering
	\includegraphics[width=0.8\linewidth]{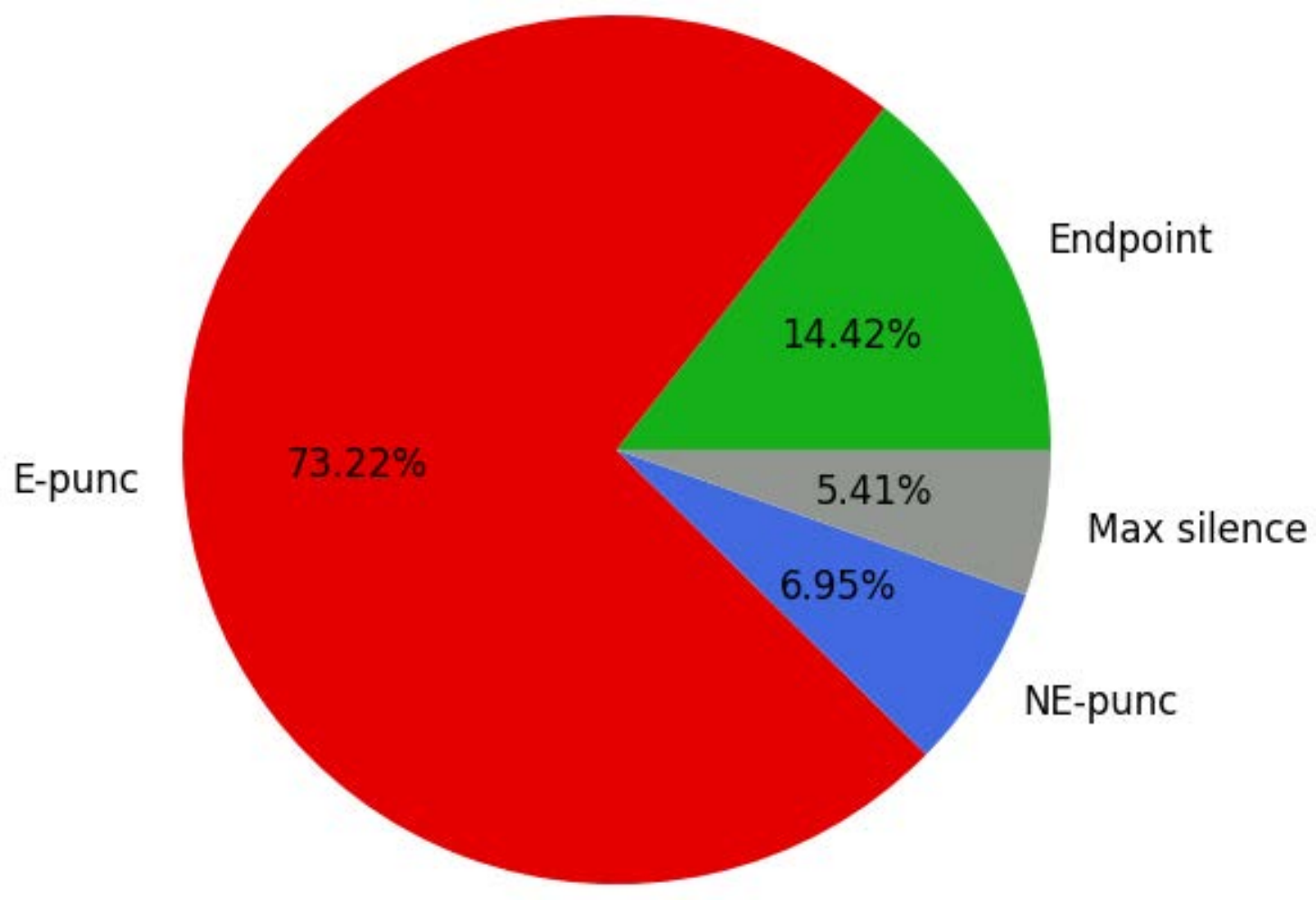}
	\caption{
	The judgment proportions of the semantic VAD.
	}
	\label{pie}
	\vspace{-0.2cm}
\end{figure}
\section{Experimental Setup}
\subsection{Dataset}
In this work, we use 325 hours of internal Mandarin speech data, obtained from various websites, such as YouTube, as training and validation sets. We use an internal Mandarin long speech test set of three different scenarios for the final evaluation, including a meeting (Test1, for 3.3 hours), a recording pen (Test2, for 1 hour), and a live webcast (Test3, for 2.1 hours). Punctuation labels per frame are obtained by forcing alignment. Table~\ref{tab:label distribution} shows the distribution of various labels in the training set.

\subsection{Evaluation metrics}
\textbf{Latency:}
Since the focus of this work is on the reduction of the tail latency of VAD segmentation during speech interaction, the latency time is taken as  one of the evaluation metrics. Specifically, the latency is measured by the average continuous tail silence duration during VAD segmentation. Since the same VAD model is used in experiments, the latency of the model itself is ignored for convenience.\\
\textbf{Character error rate (CER):}
Because the downstream task of the back-end best reflects the performance of VAD segmentation, in this work, we choose ASR as the downstream task of the back-end. After VAD segmentation, the segmented short audio will be sent to the pre-trained ASR system to obtain the corresponding transcriptions. We use pre-trained Universal ASR~\cite{abs-2010-14099} as our back-end ASR model. Since the Mandarin-speaking corpus is used for both training and evaluation, we exploit the CER as the performance measure of ASR.\\
\textbf{Detection cost function (DCF):}
The DCF is a public evaluation metric of NIST sound activity detection (SAD), which is used to measure the signal-level VAD performance in this work. In VAD segmentation,  four types of decisions might appear: True Negative (TN), True Positive (TP), False Negative (FN), and False Positive (FP). The DCF can thus be calculated as:
\begin{align}
&\text{P}_\text{miss}=\frac{\text{FN time}}{\text{TP time}+\text{FN time}}, \\
&\text{P}_\text{fa}=\frac{\text{FP time}}{\text{TN time}+\text{FP time}}, \\
&\text{DCF}=0.75*\text{P}_\text{miss}+0.25*\text{P}_\text{fa},
\end{align}
where $\text{P}_\text{miss}$ and $\text{P}_\text{fa}$  stand for the miss rate and  false alarm rate, respectively. The weights in (6) are empirically chosen. Note that a smaller DCF means a more accurate VAD segmentation.

\subsection{Model configuration}

In this work, we use the 80-dimensional log-mel filterbank (Fbank) as the
input feature. The window size is 25 ms, and the window shift
is 10 ms. Based on experience as well as tuning results on the development set, $t_\text{E}$ is set to 300ms, $t_\text{NE}$ is set to 400ms, and $t_\text{Max}$ is set to 700ms. A 6-layer contextual block conformer with 4-head of multi-head attention (MHA) is used as the speech encoder. The dimensions of MHA and feed-forward network (FFN) are set to be 256 and 512, respectively, the convolution kernel size is set to be 15, the block size is 64 (corresponding to 640 ms), and the hop size is 16 (i.e., 160 ms). In each block, the first 160 ms is the future information. For the auxiliary ASR task, a mixture of 4-layer transformer decoder and CTC~\cite{WatanabeHKHH17} is used for training. The decoder configuration keeps the same as the speech encoder with a CTC loss weight of 0.3. The  parameter amount of the speech encoder is around 5.98M. We first pre-train  the ASR task for 50 epochs, and then jointly train the punctuation prediction, ASR and VAD tasks for 50 epochs, where both $\mu$ and $\lambda$ are set to be 0.2. We use the ESPnet toolkit~\cite{watanabe2018espnet} for our experiments.
\section{Experimental Results}


First of all, we show in Table~\ref{tab:result_pipline} the comparison of the performance of traditional and semantic VAD methods on the three test sets in terms of the tail segmentation latency, back-end ASR and front-end segmentation. For a fair comparison, the same speech encoder structure as the semantic VAD is used in the traditional VAD, but the punctuation prediction is excluded. That is, only the binary classification is performed by the traditional VAD approach to distinguish speech from silence, and the semantic loss is not considered during training. It can be seen that in traditional VAD, the performance of back-end ASR becomes worse (with an average CER increase from 16.75\% to 17.94\%) as the preset tail segmentation latency decreases from 700 ms to 330 ms. The proposed semantic VAD approach can significantly reduce the average tail segmentation latency (from 700 ms to 326.92 ms) without a serious sacrifice in the back-end ASR performance (with an average CER increase from 16.75\% to 16.88\%). More importantly, the proposed method can improve the front-end segmentation performance in DCF from 10.78\% to 9.60\% due to more semantic segmentation. It is worth mentioning that in case the tail segmentation latency of traditional VAD is set to be 330 ms, it becomes equivalent to the proposed semantic VAD, while the corresponding performance of back-end ASR is much worse than that of semantic VAD. These clearly show that the proposed method can not only reduce the latency and improve the segmentation accuracy, but also preserve the back-end ASR performance.

Table~\ref{tab:ablation} presents the comparison of the proposed semantic VAD with different strategies on the three test sets in terms of the tail segmentation latency, back-end ASR and front-end segmentation. It is clear that adding the  semantic loss as an auxiliary task in training leads to better performance in all three metrics (e.g., latency from 362.95ms to 344.63ms, CER from 16.92\% to 16.75\%, DCF from 10.91\% to 9.70\% on average). This indicates that with the assistance of semantic loss, the proportion of segmentation through punctuation and endpoint becomes larger, and the semantic of the segmented sentence is more complete, which is informative to the back-end ASR module. Involving endpoint judgment in inference leads to a further reduction in  tail segmentation latency and DCF, but slightly degrades the back-end ASR performance. This is due to the fact that the endpoint is a kind of artificially defined label, which somehow increases the error in inference.

Finally, we conduct a visual analysis of the proposed semantic VAD. In Figure~\ref{visualize}, we compare the tail judgment points of traditional VAD and semantic VAD for an audio segment of Test 1. The upward and downward stems denote the judgment points of the traditional and proposed VAD methods, and the red, blue, and green stems represent the judgments of E-punc, NE-punc, and Endpoint, respectively. It is clear that the judgment points of the semantic VAD are all before those of the traditional VAD. This verifies the efficacy of the proposed method in reducing latency. Figure~\ref{pie} illustrates the proportion of various ways of obtaining tail judgment points for the semantic VAD on all test sets. It can be clearly seen that the maximum silence judgment method (i.e., traditional VAD) accounts for only 5.41\%, which indicates that our semantic VAD method can make most tail segmentation points be judged in advance. As the endpoint accounts for a relatively small proportion (14.42\%),  the artificially defined label is semantically abstract, and it is thus necessary to make a pre-judgment through punctuation prediction (e.g., with high proportions).

\section{Conclusion}
In this paper, we proposed a novelty semantic VAD method, which can synthetically determine whether tail segmentation should be performed by introducing the punctuation prediction, artificially defined endpoints, and auxiliary semantic task. Experiments on the internal dataset show that the proposed semantic approach significantly reduces the tail segmentation latency without an obvious sacrifice in the back-end ASR performance compared to the traditional VAD approach. In the future, we will investigate the application to other semantic-dependent speech tasks (e.g., speech translation, dialogue understanding) as well as the generalization to multi-task cases.


\end{document}